\documentclass[prl,amssymb,twocolumn]{revtex4}
\usepackage{graphicx}
\begin{document}

\vskip 2mm

{\bf Comment on "Universal Decoherence in Solids"}

\vskip 2mm

In a recent Letter \cite{Chudnovsky}, Chudnovsky
studied the oscillations of a quantum
particle in the double-well potential coupled to a solid. He derived the
universal lower bound on the decoherence due to phonons for the case that the
oscillation frequency $\omega_0$ is small compared to the Debye frequency
$\omega_D$. In this Comment, we show that his formula for the decoherence
rate $\Gamma$ has a limited range of validity
and is not applicable to evaluation of the width
of a low-energy optical mode considered in Ref. \cite{Chudnovsky} as an
example. This is due to unjustified use of the Fermi golden rule for
calculation of $\Gamma$. We present more general
expression for the probability of the phonon-induced transition.
For clarity, we restrict ourselves to the case of zero temperature and
assume isotropic acoustic phonons with the linear dispersion law
$\omega_{{\bf k}\lambda}=ck$, where ${\bf k}$ is the wave vector, $\lambda$
is the polarization, and $c$ is the speed of sound.

To calculate the decoherence rate for the case of a symmetric double-well
potential $U({\bf R})$, Chudnovsky makes use of the
Fermi golden rule and obtains
\begin{equation}
\Gamma=\frac{\pi m^2 X_0^2 \omega_0^2}{3\hbar\rho V}\sum_{{\bf k},\lambda}
\omega_{{\bf k}\lambda}\delta(\omega_{{\bf k}\lambda}-\omega_0)=
\frac{m^2 X_0^2 \omega_0^5}{2\pi\hbar\rho c^3}~,
\label{Gamma}
\end{equation}
where $m$ is the particle mass, $X_0$ is the half of the distance between the
degenerate minima of $U({\bf R})$, $\rho$ is the density of the crystal, and
$V$ is the normalizing volume. The value of $\hbar\omega_0$ equals to
the gap between the ground and the first
excited state of the particle. Let us recall that Eq. (\ref{Gamma}) follows
from the approximation \cite{Landau}
\begin{equation}
W_{if}(\omega,t)=
|F_{if}|^2\frac{4\sin^2\left(\frac{\omega-\omega_0}{2}t\right)}
{\hbar^2(\omega-\omega_0)^2}\approx
\frac{2\pi}{\hbar}|F_{if}|^2\delta(\hbar\omega-\hbar\omega_0)t
\label{Wif(omega,t)}
\end{equation}
for the probability $W_{if}(\omega,t)$ to find a particle in the state
$|f\rangle$ at a time $t$ if it is in the state $|i\rangle$ at $t=0$ and
interacts with the harmonic field $\hat{V}(t)=\hat{F} e^{-i\omega t}+h.c.$
Here $\hbar\omega_0=E_i-E_f$. The approximation (\ref{Wif(omega,t)})
for $W_{if}(\omega,t)$ results from the first-order perturbation theory and
is valid if (i) $W_{if}(\omega,t)<<1$ and (ii) the time $t$ is sufficiently
long, so that one can make use of the relation \cite{Landau}
$\sin^2(\varepsilon t)/\pi t\varepsilon^2\approx\delta(\varepsilon)$.

If the harmonic field $\hat{V}(t)$ is associated with a
phonon having the frequency $\omega_{{\bf k}\lambda}$, then, taking into
account that the displacements produced by the phonons with different
wave vectors are not correlated, one has for the total transition probability
\begin{equation}
W_{if}(t)=\frac{4}{\hbar^2}\sum_{{\bf k},\lambda}
|F_{if}({\bf k},\lambda)|^2
\frac{\sin^2\left(\frac{\omega_{{\bf k}\lambda}-\omega_0}{2}t\right)}
{(\omega_{{\bf k}\lambda}-\omega_0)^2}~,
\label{Wif(t)}
\end{equation}
where $F_{if}({\bf k},\lambda)$ is the matrix element for the transition
$|i\rangle\rightarrow|f\rangle$ due to the emission of a phonon
(${\bf k},\lambda)$. The form of $F_{if}({\bf k},\lambda)$ depends on the
specific nature of the states $|i\rangle$ and $|f\rangle$. For the problem
studied in Ref. \cite{Chudnovsky} it is
\begin{equation}
F_{if}({\bf k},\lambda)=mX_0\omega_0
\sqrt{\frac{\hbar\omega_{{\bf k}\lambda}}{2\rho V}}e_{\lambda}^x~,
\label{Fif}
\end{equation}
where ${\bf e}_{\lambda}$ are the unit polarization vectors. Then, making use
of the approximation (\ref{Wif(omega,t)}), one has
\begin{equation}
W_{if}^{(1)}(t)\approx\frac{2\pi}{\hbar}
\sum_{{\bf k},\lambda}|F_{if}({\bf k},\lambda)|^2
\delta(\hbar\omega_{{\bf k}\lambda}-\hbar\omega_0)t=\Gamma t~,
\label{Wif(t)approx}
\end{equation}
where $\Gamma$ is given by Eq. (\ref{Gamma}).

To quantify the applicability of the approximation (\ref{Wif(t)approx}),
let us analyze the expression (\ref{Wif(t)}) for $W_{if}(t)$. One can roughly
distinguish two contributions to $W_{if}(t)$. The first comes from the
"resonant component", i. e., from the $\delta$-function-like peak of
$\sin^2(\frac{\omega_{{\bf k}\lambda}-\omega_0}{2}t)/
(\omega_{{\bf k}\lambda}-\omega_0)^2$ as a function of $k$ at
$k_0=\omega_0/c$, with the height $t^2/4$ and the width $\sim 1/ct$.
It leads to equation (\ref{Wif(t)approx}). The second is from the
"non-resonant background" of the
phonon spectrum. At $\omega_0<<\omega_D$ and $t>>\omega_D^{-1}$ it is
\begin{equation}
W_{if}^{(2)}(t)\approx
\frac{m^2 X_0^2 \omega_0^2\omega_D^2}{4\pi^2\hbar\rho c^3}~.
\label{Wif(t)2}
\end{equation}
The Fermi golden rule
(\ref{Wif(t)approx}) for evaluation of the decoherence rate is justified if
$W_{if}^{(2)}(t)<<W_{if}^{(1)}(t)<<1$, i. e., if the "resonant component"
prevails over "non-resonant" one, and the transition probability is much
less than unity. However, this is not always the case. In the example
considered in Ref. \cite{Chudnovsky}, where an atom of mass
$m\sim 3\cdot 10^{-23}$ g oscillates at $\omega_0 \sim 10^{12}$ s$^{-1}$
in a double well with $X_0\sim 2\cdot 10^{-8}$ cm in a crystal with
$\rho\sim 5$ g/cm$^3$ and $c\sim 10^5$ cm/s,
one has $W_{if}^{(2)}(t)\sim 10$ for $\omega_D\sim 5\cdot 10^{13}$ s$^{-1}$,
i. e., the standard perturbation theory in general and the Fermi golden rule
in particular
break down. Strictly speaking, in this case the notion of a "decoherence
rate" is misleading, and one has to make use of other approaches to study the
decoherence effects. On the other hand, in the case of electron tunneling,
one has $W_{if}^{(2)}(t)\sim 3\cdot 10^{-4}$ and
$\Gamma\sim 3\cdot 10^5$ s$^{-1}$ for the same set of parameters, i. e.,
the Fermi golden rule is valid at $t>10^{-9}$ s.

Finally, it is straightforward to generalize our consideration to include
the case of an asymmetric double well and finite temperature.

\vskip 2mm

L. A. Openov

Moscow Engineering Physics Institute, Moscow 115409, Russia.

\vskip 2mm


\vskip 2mm

%\begin{references}

%\end{references}

\end{document}